\documentclass[a4paper]{article}

\usepackage{INTERSPEECH2022}
\usepackage{graphicx}
\usepackage{multirow}
\usepackage{booktabs}
\usepackage{siunitx}
\usepackage{placeins}

\title{Are disentangled representations all you need to build speaker anonymization systems? }
\name{Pierre Champion$^{1,2}$, Denis Jouvet$^1$, Anthony Larcher$^2$\thanks{This work was supported in part by the French National Research Agency under project DEEP-PRIVACY (ANR-18-CE23-0018) and Région Grand Est.}}
\address{
  $^1$Université de Lorraine, CNRS, Inria, LORIA, F-54000 Nancy, France\\
  $^2$LIUM, Le Mans Université, Avenue Olivier Messiaen, 72085 LE MANS CEDEX 9, France
}
\email{\{pierre.champion, denis.jouvet\}@inria.fr, anthony.larcher@univ-lemans.fr}

\begin{document}

\maketitle
\begin{abstract}
Speech signals contain a lot of sensitive information, such as the speaker's identity, which raises privacy concerns when speech data get collected.
Speaker anonymization aims to transform a speech signal to remove the source speaker's identity while leaving the spoken content unchanged.
Current methods perform the transformation by relying on content/speaker disentanglement and voice conversion.
Usually, an acoustic model from an automatic speech recognition system extracts the content representation while an x-vector system extracts the speaker representation.
Prior work has shown that the extracted features are not perfectly disentangled. 
This paper tackles how to improve feature disentanglement, and thus the converted anonymized speech.
We propose enhancing the disentanglement by removing speaker information from the acoustic model using vector quantization.
Evaluation done using the VoicePrivacy 2022 toolkit showed that vector quantization helps conceal the original speaker identity while maintaining utility for speech recognition.
\end{abstract}


\vspace{-2mm}
\section{Introduction}
\vspace{-1mm}

With the popularity of voice assistants, more and more connected smart speakers are being deployed in consumers' homes.
These assistants need an internet connection and centralized servers to operate. 
The user's speech is usually sent to dedicated servers for a comfortable and always-on experience.
Service providers use automatic speech recognition and natural language understanding systems to answer users' requests.
However, speech signals contain a lot of speaker-specific information, including sensitive attributes such as the speaker's gender, identity, age, feelings, emotions, etc.
Such sensitive attributes can be extracted and used as a biometric identifier or for malicious purposes such as voice spoofing \cite{asv_spoof_2017}.
This excessive and unprecedented collection of speech signals is performed to build comprehensive user profiles and construct massive datasets, which are needed to enrich and improve speech recognition and natural language understanding models.
However, this massive data collection raises serious questions about privacy.
Recent regulations, e.g., the General Data Protection Regulation (GDPR)~\cite{gdpr} in the European Union, 
emphasize the need for service providers to ensure privacy preservation and protection of personal data. 
As speech data can reflect the speaker's biological and behavioral characteristics, it is qualified as personal data \cite{nautschGDPRSpeechData2019}.

Recently, embedded speech recognition systems have been proposed to allow voice assistants to work offline.
However, the performance of these systems is still limited in unfavorable environments (i.e., noisy environments, reverberated speech, strong accents, etc.).
The alarming study conducted by \cite{fair_speech} showed significant racial disparities in the performance of widespread commercial automated speech recognition (ASR) systems.
Collecting large speech corpora representative of real users and various usage conditions is necessary to improve ASR performance inclusively. 
But this must be done while preserving user privacy, which means at least keeping the speaker's identity private.

\vspace{-1mm}
The research reported in this article deals with the problem of speaker anonymization, which aims to transform a speech signal to remove the source speaker's identity while leaving the spoken content unchanged.
This research topic has recently received renewed interest with the release of the VoicePrivacy challenge \cite{tomashenkoVoicePrivacy2020Challenge}.
The VoicePrivacy challenge (VPC) focuses on removing the speaker identity facet of a speech signal; therefore, removing personal information from the spoken content is not part of the challenge.
The baseline of the VPC \cite{fangSpeakerAnonymizationUsing2019} relies on synthesizing an anonymized speech signal from a random speaker identity, fundamental frequency, and phonetic bottleneck (ASR-BN) \cite{ppgs}, obtained from the acoustic model of an ASR system.
ASR-BN represents the articulation of speech sounds corresponding to spoken content and is supposed to be independent of the speaker's identity.
However, a significant amount of speaker information is still contained in ASR-BN \cite{adiReverseGradientNot2019,differentially_speaker_anon,mine_vq_jeps}. As they are the highest dimensional frame-level features, they restrict the speaker concealment performance of speaker anonymization systems.
This paper challenges the notion of disentanglement for the ASR-BN features.
We present a privacy-preserving ASR-BN extractor that speech synthesis systems can use to generate anonymized speech signals. 

\vspace{-1mm}
To this end, we propose to use vector quantization in an acoustic model to constrain the representation space and induce the network to suppress the speaker identity.
Vector quantization consists of the approximation of a continuous vector by another vector of the same dimension, but the latter belongs to a finite set of vectors \cite{vector_quantization}.
Vector quantization is frequently used in lossy data compression.
In our case, the compression causes the acoustic network to encode the spoken content information into a finite set of vectors.
As a result, other speaker-related information is less encoded due to a lack of encoding capacity.
The choice of the quantization dictionary size allows configuring the trade-off between utility (leave the spoken content unchanged) and privacy (speaker identity concealment).
We experimentally studied several quantization dictionary sizes to evaluate their effect on the generative capability of the speech synthesis system.
The VoicePrivacy 2022 evaluation toolkit \cite{tomashenko2020voiceprivacy_eval2022} was used to evaluate our approach empirically.

In addition to the study conducted with vector quantization, we also compare two acoustic features. Namely, the filterbanks coefficients and the wav2vec2 self-supervised speech representation \cite{wav2vec2}. 
As suggested in \cite{adiReverseGradientNot2019}, representations extracted from deeper layers in 
networks are less prone to encode speaker information.
This conclusion can be intuitively explained as deep networks contain many layers, meaning many transformations which encode the spoken content and potentially discard speaker information.
In our experiment, wav2vec2 has outperformed the filterbanks coefficients in the utility metric.

In this article, we first describe the baseline system of the VPC in Section \ref{sec:baseline}.
We then introduce the proposed model for extracting disentangled ASR-BN features in Section \ref{sec:ppg_mie} and our voice conversion system in Section \ref{sec:speech_sy}.
The experimental protocol is explained in Section \ref{sec:experiments}.
We present our experimental results in Section \ref{sec:res}.
Eventually, we draw our conclusions in Section \ref{sec:conc}.


\vspace{-2mm}
\section{Anonymization technique} \label{sec:baseline}
\vspace{-1mm}
The VoicePrivacy challenge provides two baseline systems: 
\textit{Baseline-1} that anonymizes speech utterances using x-vectors and neural waveform 
models \cite{fangSpeakerAnonymizationUsing2019} and \textit{Baseline-2} that performs 
anonymization using McAdams coefficient \cite{mcadams}. Our contributions are based on 
\textit{Baseline-1} which is referred to as the VPC 2022 baseline in this paper.
\vspace{-1mm}
\subsection{Baseline using x-vector-based anonymization}
\vspace{-1mm}
\begin{figure}[h]
  \centering
      \vspace{-2mm}
  \includegraphics[width=0.93\linewidth]{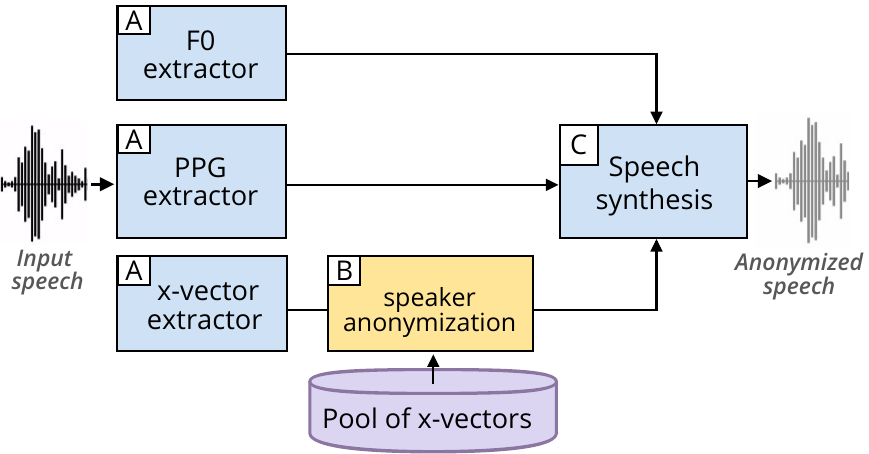}
    \vspace{-2mm}
  \caption{The speaker anonymization pipeline.}
  \label{fig:baseline.png}
  \vspace{-3mm}
\end{figure}
\noindent The central concept of the baseline system introduced in \cite{fangSpeakerAnonymizationUsing2019} is to separate speaker identity and spoken content from an input speech utterance. 
Assuming that those features can be disentangled, an anonymized speech waveform can be obtained by altering only the features that encode the speaker's identity.
The anonymization system illustrated in {Figure \ref{fig:baseline.png}} breaks down the anonymization process into three groups of modules: 
\textit{A - Feature extraction} comprises three modules that respectively extract fundamental frequency, phonetic bottleneck (ASR-BN) features from an acoustic model and the speaker's x-vector from the input signal.
Then, \textit{B - Anonymization} derives an anonymized target x-vector using knowledge gleaned from a pool of external speakers.
Finally, \textit{C - Speech synthesis} synthesizes a speech waveform from the anonymized target x-vector together with the ASR-BN features and the F0 using a neural waveform model \cite{nsf} trained with HiFi-GAN discriminators \cite{hifigan}.

\vspace{-1mm}
\subsection{The baseline ASR-BN extractor}
\vspace{-1mm}
Because of their cost functions, acoustic models used for speech recognition seek to encode spoken content information (e.g., via temporal classification of phonemes).
Thus, they are a great choice to extract phonetic posterior grams.
In the baseline of the VPC, the acoustic model used is the 17 {\scshape{tdnnf}} layers Kaldi architecture \cite{KaldiPovey,PoveyTDNN2015} and the ASR-BN is extracted from the 17th layer of the network. 
This model is trained to classify triphones with a Lattice Free Maximum Mutual Information (LF-MMI) cost functions \cite{purely_seq_lfmmi} requiring initial
alignments from a GMM model.
The dataset used for training is LibriSpeech train-other-500 and train-clean-100.

Work done in \cite{differentially_speaker_anon} has evaluated that the baseline ASR-BN extractor contains a tremendous amount of speaker-related information. Given the training pool of 921 different speakers, it is possible to identify a speaker from the ASR-BN features with 96.8\% accuracy. 
Identifying the speaker from the ASR-BN violates the disentanglement assumption of x-vector-based anonymization.

\vspace{-2mm}
\section{Proposed ASR-BN extractor method using vector quantization} \label{sec:ppg_mie}
\vspace{-1mm}
In this section, we describe the proposed vector-quantization-based ASR-BN extractor.
An acoustic model of an ASR system is also employed.
The differences between our implementation and the VPC baseline system are the following. 
Our model is trained only on train-clean-100 to reduce the computation time needed for the experiments.
Due to this aspect, we expect our model to have somewhat lower performance on non-clean speech.
The cost function used is the E2E-LF-MMI \cite{lfmmi_flat_start} criterion, allowing flat-start training without pre-training or prior alignment from a GMM model.
Our model is composed of 15 {\scshape{tdnnf}} layers, and the ASR-BN is extracted from the 13th layer.
We use a PyTorch implementation based on pkwrap \cite{pkwrap} for our model definition and training rather than Kaldi \cite{KaldiPovey}.

\vspace{-1mm}
\subsection{Vector quantization}
\vspace{-1mm}
To increase the disentanglement, we propose constraining the layer that generates the ASR-BN by using vector quantization (VQ).
Vector quantization approximates a continuous vector by another vector of the same dimension, but the latter belongs to a finite set of vectors, called prototype vectors, and is contained in a dictionary.
In the unsupervised learning task of discriminative representation via the use of auto-encoders, it has been observed that the prototype vectors learned from vector quantization primarily capture information related to the phonemes and discard some speaker information \cite{neural_disctre_vq,Unsupervised_speech_rep_vq,one-shot-vc-vector-quant}.
Similarly, the goal of applying vector quantization in an acoustic model is to induce the model to remove speaker information, as vector quantization reduces the encoding capacity of the network.
Furthermore, compared to unsupervised tasks, the cost function of an acoustic model explicitly enforces the phonetic information to be encoded.
Thus, we can apply a higher constraint by reducing the number of prototype vectors in the dictionary, which should remove even more speaker information.

\vspace{-1mm}
\subsection{VQ objective}
\vspace{-1mm}
Given the input audio sequence $s= \left(s_{1}, s_{2}, \ldots, s_{T}\right)$ of length $T$, the first {\scshape{tdnnf}} layers produces a continuous vector $h(s) = \left(h_{1}, h_{2}, \ldots, h_{J}\right)$ of length $J$ ($J < T$ due to the subsampling performed by the network) where $h_{j} \in \mathbb{R}^{D}$ for each time step $t$, and $D$ is the size of the latent representation ($D$ = 256 here).
Vector quantization takes as input the sequence of continuous vectors $h(s)$ and replaces each $h_{j} \in h(s)$ by a prototype of the dictionary $E=\left\{e_{1}, e_{2}, \ldots, e_{V}\right\}$ of size $V$, each $e_{i} \in \mathbb{R}^{D}$.
VQ transforms $h(s)$ to $q(s) = \left(q_{1}, q_{2}, \ldots, q_{J}\right)$ with:
\begin{equation}
﻿\forall j \in \left\{1, 2, \ldots, J\right\}, q_j=\underset{e_i}{\arg \min }\left\|h_j-e_{i}\right\|_{2}^{2}
\vspace{-3mm}
\end{equation}
The vector $h_{j}$ is replaced by its closest prototype vector $e_{v}$ in terms of Euclidean distance.
Since the quantization is non-differentiable (because of the $arg \min$ operation), its derivative must be approximated.
To do this, we use a \textit{straight-through estimator} \cite{strat_through_estimator} i.e.,$\frac{\partial \mathcal{L}}{\partial h(s)} \approx \frac{\partial \mathcal{L}}{\partial q(s)}$.
The prototype vectors are learned to approximate the continuous vectors which they replace by adding an auxiliary cost function:
\vspace{-1mm}
\begin{equation}
\mathcal{L}_{vq} = {\textstyle\sum_{j=1}^J} \left\|\operatorname{sg}\left[h_j\right]-q_j\right\|_{2}^{2}
\vspace{-1mm}
\end{equation}
where $\mathrm{sg}[\cdot]$ denotes the stop gradient operation, blocking the update of the weights of the {\scshape{tdnnf}} layers for this cost function (only updates the dictionary $E$).
Minimizing $\mathcal{L}_{vq}$ is a similar operation to a k-means, but applied for each minibatch during learning, the prototypes correspond to the centroids of a k-means.

Since the volume of the continuous vector space $h(s)$ is dimensionless, it can grow arbitrarily if the
dictionary $E$ does not train as fast as the {\scshape{tdnnf}}.
Adding a cost function that regularizes the {\scshape{tdnnf}} to produce continuous vector $h(s)$ close to the prototypes of $E$ is necessary so that learning does not diverge:
\vspace{-1mm}
\begin{equation}
\vspace{0mm}
\mathcal{L}_{vq\_reg} = {\textstyle\sum_{j=1}^J} \|h_j-\operatorname{sg}[q_j]\|_{2}^{2}
\vspace{-1mm}
\end{equation}

The cost function of the acoustic model can then be expressed as the sum of the MMI, quantization and regularization functions:
\vspace{-1mm}
\begin{equation}
\mathcal{L}=\mathcal{L}_{mmi}+\mathcal{L}_{vq}+\beta \mathcal{L}_{vq\_reg}
\end{equation}
where $\beta$ denotes the coefficient of the regularization factor (we used $\beta = 0.25$).
We used the learning rule based on the exponential moving average (EMA) \cite{ema_vq} to update the prototypes. 
EMA updates the dictionary $E$ independently of the optimizer, so learning is more robust to different optimizers and hyperparameters (e.g., learning rate, momentum).

\vspace{-2mm}
\subsection{Wav2vec2}
\vspace{-1mm}
In this experiment, we replaced the filterbanks coefficients used as input features for the acoustic model with wav2vec2 representation.
The model topology is adjusted, accordingly, to \cite{wav2vec2_tdnnf} we reduced the number of {\scshape{tdnnf}} layers to 9.
The ASR-BN is extracted from the 3rd layer, right before the {\scshape{tdnnf}} downsampling layer as wav2vec2 already downsampled the signal.
During training, we fine-tune the wav2vec2 model with a learning rate that is 20 times lower than the learning rate of the {\scshape{tdnnf}} layers.
We used a large wav2vec2 model pre-trained on 24.1K hours of unlabeled multilingual west Germanic speech from {V}ox{P}opuli \cite{voxpopuli}.
There is no data overlap between {V}ox{P}opuli and the data used by the VoicePrivacy evaluation plan.

\vspace{-3mm}
\section{Speech synthesis} \label{sec:speech_sy}
\vspace{-1mm}
Speaker anonymization systems usually employ voice conversion to generate a speech signal.
Given acoustic features and a target speaker representation,  voice conversion systems replace the source speaker identity with the one of the target.
In contrast to the x-vector-based speaker anonymization systems, we opted to use as speaker representation a one-hot embedding, representing the target speakers that are known and seen during training.
Furthermore, x-vector representation does not only encode speaker information \cite{prob_x-vector}. Other aspects such as the session, the speaking rate, or even non-common words are encoded.
One-hot embedding has the benefit of only encoding the speaker identifier.
Finally, we hope that given the low-dimensional one-hot speaker embedding, F0, and quantized ASR-BN, the voice conversion system will more easily convert the identity of a source speaker to another anonymized one.

\vspace{-1mm}
\subsection{F0 modification}
\vspace{-1mm}
As suggested by \cite{F0_Qian2020F0ConsistentMN}, modifying the F0 with a linear shift improves the
quality of the converted voice.
For all of our experiments, we modified the F0 mean and std to match the one of the target speaker.
Additionally, to push for the most anonymized speech, we also explored adding Additive White Gaussian Noise (AWGN) to the F0 trajectory to conceal speaker information it contains \cite{differentially_speaker_anon,F0_mod_moi,F0_trajectories}.

\vspace{-1mm}
\subsection{HiFi-GAN voice conversion}
\vspace{-1mm}
Similarly to \cite{speech_Resynthesis,language_independent_speaker_anon}, we used a HiFi-GAN-based voice conversion model to convert and generate speech.
This model achieves both high computational efficiency and audio quality.
The generator has five groups of ResBlock where multiple transposed convolutions upsample the low-frequency encoded representations of F0 features, one-hot speaker embedding, and ASR-BN, a stack of dilated residual connections are then used to increase the receptive field.
As defined in the VPC, libriTTS train-clean-100 is used to train this system.

\vspace{-2mm}
\section{Evaluation protocol} \label{sec:experiments}
\vspace{-1mm}
In contrast to the anonymization performed in the VoicePrivacy challenge, where voices are converted to random target identities on a per-speaker basis \cite{tomashenkoVoicePrivacy2020Challenge}, we convert all voices to a single target identity.
Anonymization is performed because all speakers' speech should appear to be spoken by a single identity.

	\begin{table*}[t]
	  \centering
	  \caption{Privacy and Utility scores for clean and anonymized speech on LibriSpeech test-clean and VCTK test. {\scshape{tdnnf vq} 128} indicates that the acoustic feature extractor was constrained with vector quantization and a dictionary of 128 prototypes. {\scshape{wav2vec2}} indicates that a self-supervised speech representation extractor was used instead of filterbanks. AWGN means that Additive White Gaussian Noise was added to the F0 to target a signal-to-noise ratio of 15 dB.
	  }
	  \vspace{-2mm}
	  \label{table:libri_table}
	  \begin{tabular}{
	    l@{}@{\extracolsep{0.01in}}
	    S[table-format=1.2]@{\extracolsep{0.1in}}
	    S[table-format=2.1]@{\extracolsep{0.0in}}
	    S[table-format=2.1]@{\extracolsep{0.3in}}
	    S[table-format=1.2]@{\extracolsep{0.1in}}
	    S[table-format=2.1]@{\extracolsep{0.0in}}
	    S[table-format=2.1]@{\extracolsep{0.00in}}
	    }
	    \toprule
	      \multirow{1}{0pt}{\begin{minipage}{0pt}{Dataset}\end{minipage}}
	      & \multicolumn{3}{c}{  LibriSpeech test-clean}
	      & \multicolumn{3}{c}{ VCTK test}  \\
	      \midrule
	  \multirow{2}{0pt}{\begin{minipage}{0pt}{\vspace{1mm}Method}\end{minipage}} & \multicolumn{2}{c}{ \hspace{-3mm} Privacy} & \multicolumn{1}{c}{ \hspace{-2mm} Utility} 
	 &  \multicolumn{2}{c}{ \hspace{-3mm} Privacy} & \multicolumn{1}{c}{ \hspace{-2mm} Utility} \\
	    &  \multicolumn{1}{c}{$D_{\leftrightarrow}^{\mathrm{sys}}\downarrow$} & \multicolumn{1}{c}{EER$\% \uparrow$}   &  \multicolumn{1}{c}{WER$\% \downarrow$} 
	    &  \multicolumn{1}{c}{$D_{\leftrightarrow}^{\mathrm{sys}}\downarrow$} & \multicolumn{1}{c}{EER$\% \uparrow$}   &  \multicolumn{1}{c}{WER$\% \downarrow$}\\
	    \midrule
	    Clean speech             & 0.93  & 4.1   & 4.1  & 0.93  & 2.7 & 12.8 \\ 
	    \midrule
	    VPC 2022 baseline        & 0.67  & 13.5  & 5.1  & 0.49  & 20.6  & 13.0  \\
	    \midrule
	    Ours \scshape{tdnnf no vq}     & 0.81  & 8.7   & 6.9  & 0.73  & 10.8   & 19.1 \\
	    Ours \scshape{tdnnf vq} 256 & 0.62  & 16.2 & 9.9 & 0.46  & 22.9  & 24.1  \\
	    Ours \scshape{tdnnf vq} 128 & 0.59  & 17.7 & 10.4 & 0.42  & 24.0  & 26.3  \\
	    Ours \scshape{tdnnf vq} 64 & 0.50  & 21.1  & 12.4  & 0.29  & 30.0  & 29.1  \\
	    \midrule
	    Ours \scshape{wav2vec2 tdnnf no vq}   & 0.83  & 7.7  & 3.8  & 0.69  & 12.1  & 7.8  \\
	Ours \scshape{wav2vec2 tdnnf vq} 48 & 0.57  & 17.5  & 4.5  & 0.34  & 28.0  & 10.0 \\
	Ours \scshape{wav2vec2 tdnnf vq} 48 + \scshape{f$_0$ awgn$_{15dB}$} & 0.44  & 23.4  & 4.6  & 0.12  & 40.8  & 10.3 \\
	    \bottomrule
	  \end{tabular}
	\vspace{-5mm}
	\end{table*}

\vspace{-2mm}
\subsection{Privacy evaluation}
\vspace{-1mm}
The VPC evaluation toolkit employs an automatic speaker verification (ASV) system to measure how private (speaker identity concealment) the generated speech is.
This system is an x-vector-PLDA Kaldi model and is trained on LibriSpeech train-clean-360 (same as in VPC), which was anonymized to match a single identity in our case.
This evaluation corresponds to the informed attacker scenario defined in \cite{brij_thesis}, in contrast to the semi-informed attacker scenario of the VPC.
Privacy protection is measured in terms of linkability \cite{linkability} $D_{\leftrightarrow}^{\mathrm{sys}}$ which is commonly used in biometric template protection and Equal Error Rate ($EER\%$).
The higher the $EER\%$, or the lower the $D_{\leftrightarrow}^{\mathrm{sys}}$, the better the systems are capable of anonymizing.

\vspace{-2mm}
\subsection{Utility evaluation}
\vspace{-1mm}
For the utility (spoken content recognition) evaluation, the VPC toolkit uses a Kaldi ASR system.
This model is also trained on the LibriSpeech train-clean-360 anonymized data.
The Word Error Rate ($WER\%$) metric is used. The lower the $WER\%$, the better the spoken content is preserved.

\vspace{-3mm}
\section{Results} \label{sec:res}
\vspace{-1mm}
Table \ref{table:libri_table} presents the privacy and utility performances of the proposed approaches on LibriSpeech test-clean and VCTK test datasets.
The first line shows results on clean speech, where speaker verification can be addressed with very high accuracy.

The VPC 2022 baseline system provides a strong baseline, keeping the spoken content well recognizable (absolute degradation of less than 1 $WER\%$ over clean speech) while significantly increasing privacy protection.
On the LibriSpeech dataset, privacy was increased, as the $D_{\leftrightarrow}^{\mathrm{sys}}$ metric lowered from 0.93 down to 0.67.
On the VCTK dataset, privacy was even more improved. The $D_{\leftrightarrow}^{\mathrm{sys}}$ dropped from 0.93 to 0.49.
This overall trend of seeing the VCTK dataset more anonymized than the LibriSpeech one can be explained by the dataset's nature.
LibriSpeech does not offer much variability within a single speaker due to the long recording sessions of audiobook chapters. In addition, reading speech differs from spontaneous speech, which impacts speech rate.
Those biases are captured by ASV systems \cite{rhythm_jf_odyssey,prob_x-vector}.
In the following, we will primarily focus on the VCTK results.

\noindent
Our experiment with the {\scshape{tdnnf}} ASR-BN extractor trained on filterbank without VQ shows a very high degradation of utility. In both the LibriSpeech and the VCTK datasets, the $WER\%$ increases by a large margin.
This is because the ASR-BN model was not trained with the non-clean speech of LibriSpeech train-other-500.
Even with the degradation of utility, privacy is not similarly better.
On VCTK, the $D_{\leftrightarrow}^{\mathrm{sys}}$ dropped from 0.93 to 0.73, a slight improvement but far less than the VPC baseline.
This disparity can be explained as we extracted the ASR-BN from the 13th layer while the VPC baseline extracted it from the 17th layer.
The voice conversion system might have trouble modifying the speaker identity with less refined ASR-BN.

By constraining the {\scshape{tdnnf}} network with the use of vector quantization, speaker verification performance is drastically reduced.
The number $V$ of prototypes in the quantization dictionary constrains the acoustic model.
With $V$ prototype vectors, the spoken information of the speech is compressed into a discrete dictionary space of size $V$. 
The smaller the dictionary, the more the network must find an efficient transformation to represent the spoken content information, leaving less room to encode the speaker's information.
We tried three dictionary sizes in our experiment: 256, 128, and 64.
The most anonymized speech was generated with VQ=64, where the $D_{\leftrightarrow}^{\mathrm{sys}}$ dropped from 0.73 (without VQ) to 0.29 (with VQ=64) on the VCTK dataset.
But this privacy improvement comes at a very high utility cost; the $WER\%$ raises from 19.1 to 29.1. 
The other dictionary sizes illustrate well the privacy utility trade-off \cite{privacy-utility-tradeoff} which this model suffers.
We hypothesize that the privacy improvement comes from the vector quantification layer, while the utility loss comes from the small number of layers before the quantification layer.
Constraining the network to such a few discrete vectors could be possible without significant utility loss if the network has the encoding capacity to transform the speech signal into a compressed high-level representation.

Our last experiment tested this hypothesis by using a large wav2vec2 model as a feature extractor.
Without vector quantification, our {\scshape{wav2vec2 tdnnf}} ASR-BN extractor does not significantly improve the privacy protection; the $D_{\leftrightarrow}^{\mathrm{sys}}$ on the VCTK dataset reach 0.69, far away of the 0.49 score of the VPC baseline.
Interestingly, the utility improves compared to the clean speech, the $WER\%$ drops from 4.1 to 3.8 in the LibriSpeech dataset, while in the VCTK dataset, it drops from 12.8 to 7.8.
Improvement of utility is achieved because of the {\scshape{wav2vec2}} preprocessor; the ASR-BN is more precise because of the network depth and amount of training data that the {\scshape{wav2vec2}} was trained on.
Applying voice conversion on precise ASR-BN normalizes the speech signal allowing the ASR system to better recognize the spoken content.

Applying a high vector quantification constraint on this {\scshape{wav2vec2 tdnnf}} model shows the approach's potential.
With a very small dictionary size of 48 prototypes, privacy is improved in comparison to the VPC baseline; the $D_{\leftrightarrow}^{\mathrm{sys}}$ on the VCTK dataset reaches 0.34 while also improving the utility with 10.0 of $WER\%$.
To push privacy preservation to the extreme, we added white Gaussian noise to the F0 trajectory to hide the speaker information that it contained. 
This modification increased the privacy protection, as the $D_{\leftrightarrow}^{\mathrm{sys}}$ on the VCTK dataset plummeted down to 0.12 while keeping a very high utility with 10.3 of $WER\%$, similar behavior can be observed in the LibriSpeech dataset.

\vspace{-3mm}
\section{Conclusion} \label{sec:conc}
\vspace{-1mm}
This paper challenged the notion of feature disentanglement at the ASR-BN, F0, and speaker representation levels.
We proposed to use a vector-quantized-based ASR-BN feature extractor as disentangled acoustic representation. 
Experiments on the VPC 2022 datasets demonstrated that our proposed speaker anonymization method based on extracting ASR-BN from a deep acoustic model constrained with vector quantification generates anonymized speech which greatly protects users' privacy while improving the utility.
\textbf{However}, the dataset of the VoicePrivacy uses clean speech, which is favorable for this approach, under noisier environments, the utility largely decreases.
We also emphasize, that if the F0 is used for anonymizing the voices of a small database, a modification of it needs to be done.
While noise-based modification improves privacy, it degrades human intelligibility and naturalness.
Promising results can be achievable by quantizing the F0.
Finally, to answer the question of the title, we believe that yes, disentangled representations are all you need, however, their extraction remains a challenging task, especially under noisy, weakly labeled, multilingual conditions.
Maybe except for the speaker representation, where a simple one-hot encoding is all you need.
Live demos of the anonymization system, pre-trained models, and source code are available at: https://colab.research.google.com/github/deep-privacy/SA-toolkit/blob/master/SA-colab.ipynb 


\bibliographystyle{IEEEtran}

\bibliography{mybib}


\end{document}